\documentclass[aps,pra,reprint,superscriptaddress,twocolumn]{revtex4-2}
\usepackage[utf8]{inputenc}
\usepackage[T1]{fontenc}
\usepackage{amsmath,amssymb}
\usepackage{float}
\usepackage{graphicx}
\usepackage{lineno}
\usepackage{blindtext}
\usepackage{xcolor}
\usepackage{tikz}
\usepackage{siunitx}

\begin{document}
\title{Optimal Superpositions for Particle Detection via Quantum Phase}

\author{Eva Kilian}
%\email{eva.kilian.18@ucl.ac.uk}
\affiliation{%
 Department of Physics \& Astronomy, University College London, London WC1E 6BT, UK 
}
 
\author{Marko Toro\v{s}}
\affiliation{%
School of Physics \& Astronomy, University of Glasgow, Glasgow, G12 8QQ, UK
}
\author{P.F. Barker}
\author{Sougato Bose}
\affiliation{%
 Department of Physics \& Astronomy, University College London, London WC1E 6BT, UK 
}

%\date{\today}
\newcommand{\mt}[1]{{\color{blue}MT: #1}}

\begin{abstract}
   Exploiting quantum mechanics for sensing offers unprecedented possibilities. State of the art proposals for novel quantum sensors often rely on the creation of large superpositions and generally detect a field.  However, what is the optimal superposition size for detecting an incident particle (or an incident stream of particles) from a specific direction? This question is nontrivial as, in general, this incident particle will scatter off with varied momenta, imparting varied recoils to the sensor, resulting in decoherence rather than a well defined measurable phase.  By considering scattering interactions of directional particulate environments with a system in a quantum superposition, we find that there is an ``optimal superposition'' size for measuring incoming particles via a relative phase. As a consequence of the anisotropy of the environment, we observe a novel feature in the limiting behaviour of the real and imaginary parts of the system's density matrix, linking the optimality of the superposition size to the wavelength of the scatterer.  
\end{abstract}

\maketitle

\begin{figure*}[ht]
     \centering
     \includegraphics[width=\textwidth]{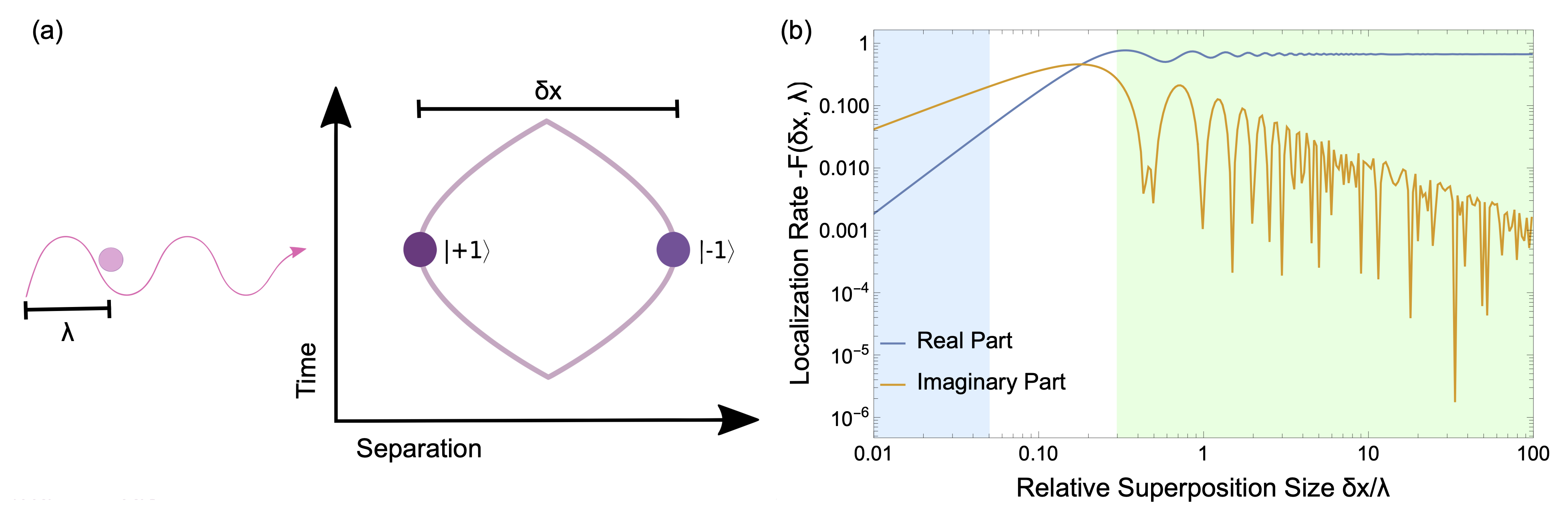}
     \caption{(a) Basic schematics of a particle (pink) scattering from a quantum object (purple spheres) in a Stern-Gerlach type interferometric experiment, where the sensor's spin-state dependent evolution is exploited to split and recombine the motional wavepacket (purple line). The incoming particle is sensed through its momentum transfer to the quantum object, which manifests in the appearence of a relative phase between superposed components. (b) Short- and long-wavelength regimes of quantum state evolution for Thompson scattering. Regions shaded indicate the long wavelength regime in blue, the intermediate region in white and a green shaded region that includes the short-wavelength regime. The quantity $\lambda$ is the incoming particle's wavelength, $\delta x$ is the size of the superposition, Re and Im refer to the behaviour of the real and imaginary parts of the localization rate's contribution to the  off-diagonal elements of the system's density matrix. The real part (blue line) is typically used to quantifiy decoherence. The imaginary part (orange) vanishes in an isotropic situation, where the particles are coming randomly from all directions.  When we have a directed (non-isotropic) source such as a stream of particles however, then the imaginary part can be used for sensing particles through relative phases, particularly in regimes where the decoherence contribution (blue line) is suppressed. The Goldilocks region for the superposition size for optimal sensing is in the intermediate white-shaded region.}
    \label{fig:main_figure}
\end{figure*}

%\section{Introduction}
Quantum sensing with matter-wave interferometers has prompted the development of a variety of commercial technologies and experiments, offering some of the most precise sensors. State of the art experiments and proposals encompass research in the areas of metrology~\cite{pezze_quantum_2018}, gravimetry~\cite{asenbaum2017phase,scala_matter-wave_2013,   wan_free_2016,qvarfort_gravimetry_2018,marshman_mesoscopic_2020,torovs2021relative,wu2023quantum}, geophysics~\cite{bidel_absolute_2018,stray_quantum_2022}, quantum foundational principles~\cite{bose_spin_2017,marletto_gravitationally_2017,bose2023entanglement,biswas2022gravitational,margalit2021realization,yi2021massive}, and sensing for fundamental physics \cite{dimopoulos2009gravitational,chaibi2016low,marshman_mesoscopic_2020,geraci2016sensitivity,riedel_direct_2013,bateman2015existence,riedel2017decoherence,kilian_requirements_2023}. While larger quantum objects as a sensor, such as a nanoparticle, to date have typically been prepared in a near-classical or Gaussian (which can be somewhat quantum in the sense of being squeezed) initial state \cite{afek_coherent_2022,burd_quantum_2019,carney2021mechanical}, the full potential of quantum mechanics becomes apparent when non-Gaussian states, such as a quantum superposition of two distinct states, or a state during interferometry, are utilized. Optimization of the experimental setup and parameters in such cases in order to extract exquisitely weak signals is of utmost importance.\\
The sensing of potentials, such as a linear potential generated by electrostatic fields, or the gravitational potential near earth, often necessitates the realisation of quantum superposition states with large spatial separation $\delta x$ between the superposed components, since the accumulated phase $\Delta\phi$ increases with increasing separation. For example, an object of mass $m$ held in a quantum superposition of localized states separated vertically by $\delta x$ for a time interval $\tau$ near earth's surface acquires the celebrated Collela Overhauser Werner (COW) phase of $\Delta \phi \sim m g \delta x \tau/\hbar$ \cite{colella1975observation,scala_matter-wave_2013,wan_free_2016}, while the phase due to the curvature of a proximal mass is $\propto (\delta x)^2$ \cite{torovs2021relative}. For a dynamical monochromatic classical field of wavenumber magnitude $k$, again, while phase $k\delta x$ is defined modulo $2\pi$, it surely does not harm the coherence of the superposition to have $\delta x > 1/k$\cite{geraci2016sensitivity}.  At the other extreme is the detection of particulate matter interacting with the sensing system via a coupling term. This is, however, phenomenologically different as it cannot be correctly approximated by a classical field, unless in the macroscopic limit of a very large number of particles in a coherent state. In this limit, the focus has been on detection of the particulate source via the decoherence of a quantum superposition, manifested as loss of interference or fringe visibility, as the particles scatter off the sensor mass, typically imparting random momenta \cite{riedel_direct_2013,bateman2015existence,fragolino2023decoherence}. Thus, the measurement of a coherent phase is typically not associated with the detection of particles. As far as current understanding goes, the intuition is that if fields are concerned, $\delta x$ higher is typically better, and surely not harmful, while for particles, decoherence would be the prime signature.\\
In this work, we show that {\em neither} of the above intuitions are correct for particulate matter incident from a given direction: We find that there exists an {\em optimal} superposition size for quantum sensing in scattering experiments, depending on the characteristics of the environmental source. This arises due to a competition between a coherent phase contribution and a decoherence contribution. To illustrate this effect, we consider the basic blueprint consisting of an incoming particle (the signal) that scatters from a massive quantum system placed in a spatial superposition (the sensor) as is schematically represented in Fig. \ref{fig:main_figure} (a). Working within the framework of open quantum systems, we compute the effects arising from the interaction of the system with a directional particulate matter environment and discuss in how far the superposition size impacts the accumulation of the phase. Contrary to expectations, reading the phase imparted due to scattering in the presence of decoherence induced by the same scattering may not be optimized at a trivial point. Aside the above fundamental point, we also present an application in single photon and atom detection. We conclude by discussing the implications of our finding for present-day sensing experiments aimed toward capturing signatures of other/general scattering particles.

\section{Scattering Master Equation}

The appearance of the classical from the quantum is the cause of long-standing debate and a key element of all interpretations of quantum mechanics. The process of environmental decoherence offers one mechanism for this quantum-to-classical transition, through which the suppression of quantumness following measurements performed on a sensing system initially in superposition of several eigenstates may be described. In this context, the interaction of a superposed quantum object with gaseous particles and 
photons will constitute a measurement, since it involves the transfer of positional information into the environment. Following the work of Joos and Zeh~\cite{joos_emergence_1985}, the authors of~\cite{gallis_environmental_1990} have extended and generalised the mathematical model reflecting the localizing mechanism under the critical assumption that the scattering of the (weakly coupled) environment does not significantly disturb the sensor.\\
Concretely, in a non-relativistic theory, the reduced density matrix of a system interacting with a particulate environment through scattering~\cite{gallis_environmental_1990} is governed by the following master equation, which includes the corrective factor derived in~\cite{PhysRevA.68.012105,PhysRevLett.90.160401}, 

\begin{align}
\label{eq:GF}
    \frac{d\hat{\rho}_S(\bold{x},\bold{x}')}{dt}&=\frac{1}{i\hbar}\langle \bold{x}|[\hat{H}_S,\rho_S]|\bold{x}'\rangle\\ \nonumber
    &-F(\bold{x}-\bold{x}')\hat{\rho}_S(\bold{x},\bold{x}'),
\end{align}
where $\hat{H}_S$ is the Hamiltonian that describes the unitary evolution of the sensor, $\hat{\rho}_S$ the density matrix of the sensing system and
\begin{align}
\label{eq:localization_rate_1}
    F(\bold{x}-\bold{x}')&=\int dqn(q)v(q)\int \frac{d\Omega d\Omega'}{4\pi}p(\Omega,\Omega')\\ \nonumber
    &\times(1-e^{i(\bold{q}-\bold{q}')(\bold{x}-\bold{x}')})|f(\bold{q},\bold{q}')|^2. 
\end{align}
This result is derived through a careful perturbative treatment of the scattering interaction. The function $p(\Omega,\Omega')$ is a normalized probability density with incoming (outgoing) scattering angle $\Omega$ ($\Omega'$), and where $p(\Omega,\Omega')=1$ is the typical scenario where scatterers impart momentum from all directions. 
The quantities $n(q)$ and $v(q)$ refer to the number density and speed of particles with wavenumber $q$, the latter of which is closely related to their wavelength $\lambda=\frac{2\pi}{q}$ and momentum $\bold{p}=\hbar\bold{q}$, with $\bold{q}$ being the wave vector. The scattering amplitude of the relevant interaction process is denoted as $f(\bold{q},\bold{q}')$, with $\bold{q}$ and $\bold{q}'$ labelling incoming and outgoing wavevectors respectively. $F(\bold{x}-\bold{x}')$ is the localization rate. If it is real-valued, the sensing system exhibits a loss of coherence over time, while an imaginary contribution manifests in the appearance of a phase $e^{i(\bold{q}-\bold{q}')(\bold{x}-\bold{x}')}$. Differences in the phases arising at $\bold{x}$ and $\bold{x}'$ can be measured and exploited in sensing with quantum systems.\\
For an incoming particle of wavelength $\lambda$, it is useful to divide the system to two regimes of interest to investigate in order to describe the behaviour of the sensing system.\\
In the long-wavelength regime, where $\lambda \gg \delta x$, with $\delta x = |\bold{x}-\bold{x}'|$, the phase term in Eq. (\ref{eq:localization_rate_1}) becomes small enough to warrant an approximate treatment of the exponential function by means of a Taylor-expansion of the argument. Calculating the expansion up to second order, a quadratic dependency of the localization rate on the superposition size $F(\bold{x}-\bold{x}')\propto \frac{1}{2}q^2(\bold{\hat{n}}-\bold{\hat{n}}')\cdot(\bold{x}-\bold{x}')^2$ is revealed. Assuming isotropy of the environmental source, the linear term in the expansion averages to zero due to the integration over terms involving the product of an even and odd function in directions $\bold{\hat{n}}$, $\bold{\hat{n}}'$.\\
Master equations of the form (\ref{eq:GF}) can be mapped to equations of the Lindblad-type in position representation and for $\hat{L}_k$ corresponding to a physical observable, namely that of the position operator $\hat{L}_k=\bold{\hat{x}}$, the equation that governs the evolution of the system can be expressed as

 \begin{align}
 \label{eq:Lindbladform}
\frac{d\hat{\rho}_S(\bold{x},\bold{x'},t)}{dt}=-\frac{\kappa}{2}(\bold{x}-\bold{x}')^2\hat{\rho}_S(\bold{x},\bold{x'},t),  
\end{align}
assuming the individual evolution of the sensing system's density matrix to be negligible. The quantity $\kappa$ incorporates information contained in the localization rate $F(\bold{x}-\bold{x}')$ as written in Eq. (\ref{eq:localization_rate_1}). \\
In the short-wavelength regime, the exponential function in Eq. (\ref{eq:localization_rate_1}) oscillates rapidly and hence, quickly averages out when performing the integrals. Eq. (\ref{eq:localization_rate_1}), expressed in the form of Eq. (\ref{eq:Lindbladform}), tends to

\begin{align}
\label{eq:SWL}
    \frac{d\hat{\rho}_S(\bold{x},\bold{x'},t)}{dt}=-\frac{\Gamma}{2}(1-\delta_{\bold{x},\bold{x'}})\hat{\rho}_S(\bold{x},\bold{x'},t).  
\end{align}
 where, just for a qualitative understanding of the behaviour, we have taken a discrete set of $\bold{x}$ values and $\delta_{\bold{x},\bold{x'}}$ is a Kronecker delta. The quantity $\Gamma$ incorporates information contained in the localization rate $F(\bold{x}-\bold{x}')$.
 If a given environment is not isotropic and the scattering particles are instead impinging from a specific direction, the limiting behaviours reveal the emergence of an optimal superposition size where the detection of the particle is also from the phase imparted. In order to demonstrate this effect, we now resort to a combination of analytic approximations and exact numerics of the reduced system density matrix for explicit examples.

\section{On the Emergence of an Optimal Superposition Size for Particle Detection}

To illustrate the emergence of an optimal superposition size, let us focus on two cases of a well-known example from the literature~\cite{gallis_environmental_1990}, where a special form of the differential cross-section is taken to be

\begin{equation}
\label{eq:differential_crosssection}
    |f(\bold{q},\bold{q}')|^2=gq^j\frac{1}{2}\Bigl(1+\left|\frac{\bold{q}\bold{q}'}{q^2}\right|^2\Bigr).
\end{equation}
For the expansion order $j=0$ and $g=r_e^2$ with $r_e$ the square of the classical electron radius, one recovers a differential cross-section for Thompson scattering, while values of $j=4$ and $g=a^6|\frac{\epsilon-1}{\epsilon+1}|^2$ with $a$ as the scatterer's radius and $\epsilon$ being the dielectric constant lead to a description of Rayleigh scattering.\\
For all practical purposes, let us assume the directional source of particles to travel along the z-Axis and the superposition to be oriented equally along z. In the long wavelength-limit, we are again able to expand the small exponent in Eq. (\ref{eq:localization_rate_1}) in orders of $\delta x$.  This assumption is reflected in the choice of $p(\Omega,\Omega')=\delta(\theta)\delta(\phi)/\sin{\theta}$ for a spherical coordinate system. Further, we select the coordinates of our incoming and outgoing particle wave vectors to be $\bold{q}=q(0,0,1)$, $\bold{q}'=q(\cos\varphi'\sin\theta',\sin\varphi'\sin\theta',\cos\theta')$, where we notably keep the magnitude of the particle's wave vector and hence momentum unchanged, which is a valid approximation for negligible momentum transfers.\\ If $j=0$, calculating the Taylor expansion up to second order in $\delta x$ and performing the subsequent angular integration results in the following terms of the localization rate
\begin{align}\label{eq:m=0locrate}
    F(\bold{x}-\bold{x}')&=\int dq n(q)v(q)g\Bigl(-\frac{2}{3}i q\delta x\\ \nonumber
    &+\frac{7}{15} q^2\delta x^2 +\mathcal{O}(\delta x^3)\Bigr).
\end{align}
Similarly, we obtain a barely modified equation for \mbox{$j=4$}
\begin{align}
    F(\bold{x}-\bold{x}')&=\int dq n(q)v(q)gq^4\Bigl(-\frac{2}{3}i q\delta x\\ \nonumber
    &+\frac{7}{15} q^2\delta x^2 +\mathcal{O}(\delta x^3)\Bigr).
\end{align}
For both the above cases, we observe the emergence of an imaginary linear (Hamiltonian) term in the master equation, sometimes known as the effective index of refraction\cite{PhysRevLett.74.1043,vacchini_quantum_2009}, describing the evolution of the reduced density matrix, reported in Tab. \ref{tab:Limits}, showing the possibility of the detection of the particles by a phase. Importantly, this behavior appears to be independent of the exact form of the differential scattering cross section (i.e., it can work in more generality, even beyond the above two cases), surfacing merely due to the directional momentum impartment of the incoming scatterers.\\

\begin{table}[h!]
    \centering
    \begin{tabular}{||c|c|c||}
    \hline
        & real part & imaginary part\\
         \hline\hline
     long-wavelength limit &$\propto \delta x^2$ &$\propto\delta x$\\
       short-wavelength limit   &$\Gamma= \text{constant}$&$\rightarrow$ 0\\
       \hline
    \end{tabular}
    \caption{Limiting behaviour of the off-diagonal elements of the sensing system's density matrix. The behaviour of the real part follows the theoretically predicted trends as observed already in random uniform scattering, see Eq. (\ref{eq:Lindbladform}) and Eq. (\ref{eq:SWL}). However, for a non-uniform environment, the imaginary part ($\mathcal{O}(2)$) exhibits a drastically different behaviour and is non-vanishing.}
    \label{tab:Limits}
\end{table}
The limit of small wavelengths is more difficult to treat analytically due to the rapid oscillatory behaviour of the integrand. We can however significantly reduce the complexity of the problem by assuming the specific geometry described at the beginning of this section. We then subsequently employ the Jacobi-Anger expansion, noting the cosine appearing in the exponent. Through this type of expansion, the trigonometric function in our exponential is expressed in the basis of its cylindrical harmonics via the relation $e^{iz\cos{\theta}}=J_0+2\sum_{n=1}^\infty i^nJ_n(z)\cos{n\theta}$. This expansion enables us to numerically evaluate the real and the imaginary parts of the localization rate, as well as the signals in various phase measurements.\\
In what follows, in order to simplify our presentation when making plots, we assume a very narrow distribution of incoming momenta and hence wavenumber magnitude, namely given by $\delta(q-q_0)$. Figure \ref{fig:main_figure} (b) displays the trend of the real and imaginary parts for multiple values of the ratio $\frac{\delta x}{\lambda}$, where the factor $g=r_e^2$, effectively scaling the plot's y-axis, has been neglected and the momentum integration has been performed under the assumption of a delta-function in incident momentum, which implies a specific $q$ scattering. In the small $\delta x/\lambda$ limit, the full numerics clearly match with the polynomially increasing behaviour of the imaginary (linear) and real (quadratic) parts as given by Eq.(\ref{eq:m=0locrate}). Moreover, the visible decay of the imaginary phase contribution to zero and the saturation of the real part also confirm the expected limiting behaviour for large values of $\frac{\delta x}{\lambda}$ in the short wavelength-limit. However, the way in which these limits are reached differs drastically from the case of random uniform scattering, where the exponential contribution habitually averages out due to rapid oscillations (that is the standard setting of decoherence due to scatterers \cite{joos_emergence_1985,gallis_environmental_1990})\\
The second example for $j=4$ reveals a similar trend, though the resulting localization function is again scaled by an additional factor of $q^4$ and the coupling $g$ is dependent on the dielectric constant of the sensing material. Employing previous calculational methods will therefore lead to the same qualitative observation, namely the emergence of an optimal "Goldilocks zone" in relation to the accumulation of phase as depicted in Fig. \ref{fig:main_figure}, where the zone is shaded in white and the phase imparted is optimal for $\delta x \sim 0.2 \lambda$.

\section{Experimental Signature}

The field of matter-wave interferometry offers a catalogue of schemes that enable the extraction of the phase contribution arising due to a scattering interaction. A popular approach is founded in the Stern-Gerlach interferometry of spin-mechanical systems~\cite{bose_spin_2017,marshman_constructing_2022}, where the magnetic manipulation of a mesoscopic test-mass with an embedded spin, such as levitated diamond with a nitrogen-vacancy-center, is used. After the initialization of the sensing system in a center-of-mass (COM) motional state $|c\rangle$ and a superposition of (its embedded) spin states $\frac{1}{\sqrt{2}}\bigl(|+1\rangle+|-1\rangle\bigr)$, the system is allowed to evolve. This evolution is spin-state dependent and leads to a spatial splitting of the COM, resulting in a combined quantum state of the form $|\psi(t)\rangle=\frac{1}{\sqrt{2}}\bigl(|c_{s=+1}(t)\rangle|+1\rangle+|c_{s=-1}(t)\rangle|-1\rangle\bigr)$. The difference $\phi$ in distinct phases arising between the components of the superposition due to a scattering interaction can be measured upon completion of the interferometer purely from the spin states. In the ideal case of no decoherence, it becomes $\frac{1}{\sqrt{2}}\bigl(|+1\rangle+e^{i\phi}|-1\rangle\bigr)$. For an illustration of this scenario, see Fig. $\ref{fig:main_figure}$ (a). In presence of a decoherence term $A\sim e^{-\Gamma t}$ the density matrix of the spin at the end of the interferometry becomes
\begin{align}\rho=\frac{1}{2}\begin{pmatrix}a&Ae^{i\phi}\\Ae^{-i\phi}&b\end{pmatrix}.
\end{align} 
As the elements of the density matrix are going to be estimated from probabilities of various measurements, it is clear that the exponentially decaying decoherence term $A$ is less good as an estimator than $Ae^{\pm i\phi}$ when the term $A\sim 1$ (varying $\phi$, say by controlling $\delta x$ can give oscillations in probabilities raging from 0 to 1). Thus the phase effect found by us here from a directional source of particles presents an important method to detect them in comparison to the decoherence they produce.

A method of extracting phase-differences arising between the off-diagonal components is to apply $\pi/2$-phase and Hadamard gate transformations $S$ and $H$ to the quantum state, which effectively results in a projection of the phases onto the diagonal elements of the density matrix 

\begin{align}
    \rho_{f}&=HS\rho S^\dagger H\\\nonumber
    &=\frac{1}{4}\begin{pmatrix}a+b+2A\sin{\phi}&a-b+2iA\cos{\phi}\\a-b-2iA\cos{\phi}&a+b-2A\sin{\phi}\end{pmatrix}.
\end{align}

\begin{figure*}[t]
    \centering
    \includegraphics[width=\textwidth]{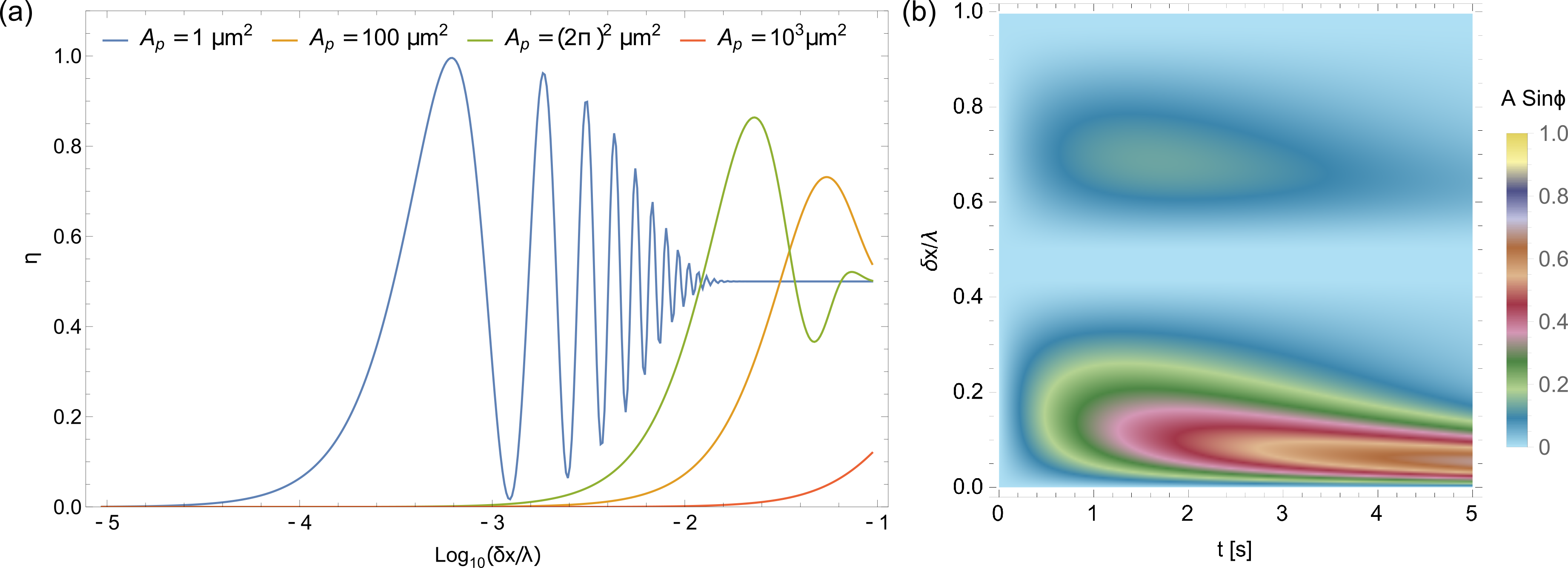}
\caption{(a) Quantum efficiency $\eta$, as defined in Eq. (\ref{eq:efficiency}), for Rayleigh scattering of single photons with $\lambda=1064$ nm on a 0.1 micron-sized sphere, assuming different spatial photon profiles $A_p$ denoting the transverse area of the incoming photon assuming an estimated flux of $10^6$ photons per area $A_p$ per second. (b) Measurable signal (depth coloration) for varying ratios of $\delta x/\lambda$ over a time interval $t=[0,5]$ s, assuming the illustrative example of Thompson-type scattering with the exponentiation index for $q^j$ in Eq. (\ref{eq:differential_crosssection}) as $j=0$ and taking the momentum distribution to be a delta function (definite momentum, say, around $q=q_0=2\pi/\lambda$) and setting the incident flux of the particles to be such that $\int g n(q) v(q) dq\approx 1$. The orange colored region in the plot indicates the maximization of the phase signature in the long-wavelength limit for the lower island, which is also indicated by the non-vanishing first order contribution in Eq. (\ref{eq:m=0locrate}). Depending on the narrowness of the momentum distribution, the depicted behaviour of the signal should exhibit similar features in the case of atom scattering.}
\label{fig:phaseplot_eff}
\end{figure*}
Subtracting the (diagonals) probabilities 
\begin{align}
\rho_{f,11}-\rho_{f,22}=A\sin{\phi},
\end{align}
hence relates to the sine of the accumulated phase $\phi$.  This experimental signature is plotted in Fig. \ref{fig:phaseplot_eff} (b) for the normalized initial state where $\rho_{11}=\rho_{22}=\rho_{12}=\rho_{21}=1/2$ and evolved final states with varying values $\phi=2\pi\delta x/\lambda$, at times $t=[0,5]$s. For the sake of simplicity and to concentrate on qualitative aspects of the core issue, i.e., the results of angular averaging over outgoing momenta, we have not chosen a specific distribution for the number density and speed of the particles and taken the momentum distribution to be a delta function (definite wavenumber around $q=q_0$) and set the incident flux of the particles to be such that $\int g n(q) v(q) dq\approx 1$, effectively plotting the contributions from $F(x-x')$ resulting from the angular integrations. Specific values for these quantities and subsequent integration over the wavenumbers will lead to a shifted optimal range for the relative superposition size $\delta x/\lambda$. Although we do not strictly define the notion of optimality, it should be implicitly clear that values of the relative superposition size leading to a phase contribution of $\mathcal{O}(1)$ are considered as such. Benchmarking the range of suitable values for $\delta x/\lambda$ to, say, $\sin(\phi)=0.95$ is one possible way of defining the window of optimality. For larger or smaller fluxes, the time needed for $\mathcal{O}(1)$ will be appropriately scaled.

\subsection{Single Photon Detection}

To give a concrete example of an experimental application, we analyse the potential benefit of this novel effect for the detection of spatially shaped single-photons, such as those which may be emitted from a quantum dot source~\cite{rakher_simultaneous_2011,hanschke_quantum_2018}. With the aim of operating our sensing system as a "click" detector, we introduce what we term the detection efficiency as follows

\begin{align}
\label{eq:efficiency}
    \eta&=\langle -|\rho_{in}(t=1\text{s})|-\rangle\\ \nonumber
    &=\frac{1}{2}(1-(\rho_{12}(t=1\text{s})+\rho_{21}(t=1\text{s}))).
\end{align}

The efficiency $\eta$ quantifies the distinguishability of a system initially prepared in a superposition state $\rho_{in}=|+\rangle\langle +|$ from its final state. If the scattering of a photon results in a $\pi$ phase shift, projecting the initial quantum state into the orthogonal state $|-\rangle$, the efficiency approaches its maximum. Figure \ref{fig:phaseplot_eff} (a) shows the crucial dependency of $\eta$ on the choice of $\delta x/\lambda$ for different magnitudes of spatial photon profiles $A_p$. The single photon transverse area corresponds to the inverse of $n(q)v(q)\tau$ with $\tau$ set to be 1s and $n(q)v(q)\tau\sim \frac{10^6}{A_p}$. Whereas certain choices of the superposition size will be suitable for operating our sensing system as a single photon detector, others will result in phase shifts that render the system insensitive to the signal, implying that the superposition size can be used for wavelength selection. It can be seen in Fig. \ref{fig:phaseplot_eff}(a) that sensing will be possible within the peaked regions of relatively broad bands.\\  

\subsection{Detection of Single Atomic Ions}

Rutherford scattering of an atomic ion on a nanoparticle of radius $1\mu$m is described via the differential cross section

\begin{align}
\label{eq:atoms}
    |f(\bold{q},\bold{q}')|^2=\frac{m^2}{\hbar^4 q^4}\frac{(ZZ'e^2)^2}{(4\pi\epsilon_0)^2}\Bigl(1+\left|\frac{\bold{q}\bold{q}'}{q^2}\right|^2\Bigr),
\end{align}

where $Z e$ and $Z' e$ are the charges of the atom and nanoparticle, $m=10^{-25}$kg is the mass of a heavy atom and $\epsilon_0$ the vacuum permittivity. We assume an atomic momentum determined via the relation $k_BT/2=\hbar^2\bold{q}^2/2m$, at a temperature $T=100$K. Expressing the momentum and inserting into Eq. (\ref{eq:atoms}) yields

\begin{align}
|f(\bold{q},\bold{q}')|^2=10^{-14}Z'^2\Bigl(1+\left|\frac{\bold{q}\bold{q}'}{q^2}\right|^2\Bigr),
\end{align}

where the atom's charge number $Z=1$ has been assumed. We hence propose that a micron sized nanoparticle is able to detect about one atomic ion interacting with it per second if the atom has a wavefunction cross section on the order of $10^4$nm$^2$ for small values of $Z'$ and the atomic flux (corresponding to $n(q)v(q)$) is on the order of $\sim 10^{-4} \text{s}^{-1}\text{nm}^{-2}$). 

\section{Summary}
Our observations are of critical relevance to experiments where a stream of incoming particles scattering from a superposition has a unique or preferred direction and the environment cannot be treated in the fashion of an isotropic bath. Conversely, we expect that a similar observation can be made for a superposed object that is not held in place, but instead propagating with a given velocity with respect to the environment, such as a crystal with horizontal velocity in a motional superposition state moving through a gas of particles in the lab frame.\\
Considering two limiting regimes for the wavelength of an incoming scatterer interacting with a quantum sensor, we have numerically shown that the imaginary contribution arising due to the interaction is, in specific scenarios, non-vanishing and we have provided a strong argument that the behaviour is likely universal. The described relative phase may be used for the detection of weak environmental signatures.\\ Moreover, we have observed the emergence of an optimal parameter-choice for the superposition size $\delta x$ when it comes to measuring special types of particulate environments and identified a Goldilocks-zone. We have also shown that the superposition sensor is capable of detecting single photons and single charged atoms with appropriate settings.\\
Our findings will doubtlessly result in improvements of state-of-the-art quantum sensors and may be utilized to enhance signals which are typically difficult to capture. Several emergent experiments and technologies~\cite{scala_matter-wave_2013,marshman_mesoscopic_2020,bateman2015existence}, especially in the context of quantum gravimetry~\cite{rademacher_quantum_2020}, rely on the acceleration of the quantum mechanical sensor. Any such setup will be influenced by non-isotropic sources, such as the directional scattering of a stream of particles, interacting with the sensing apparatus. We therefore want to emphasize the importance of the choice of the superposition size in relation to phase contributions arising through these directional effects.% In turn, we also highlight the potential of our findings to be exploited for the purpose of intentional enhancement of a particular anisotropic signal. 

\acknowledgements
EK would like to acknowledge support from the Engineering and Physical Sciences Research Council (grant number EP/L015242/1) and an EPSRC Doctoral Prize Fellowship Grant (EP/W524335/1). MT would like to acknowledge funding by the Leverhulme Trust (RPG-2020-197). SB would like to acknowledge EPSRC grants (EP/N031105/1 and EP/S000267/1) and grant ST/W006227/1. PFB acknowledges funding from the EPSRC Grant No. EP/W029626/1.

\bibliographystyle{apsrev4-1} % Tell bibtex which bibliography style to use

\bibliography{references.bib}

\end{document}